\documentstyle[12pt,psfig,draft]{nature-dbf}

\newcommand{\Msun}{M_{\odot}}
\newcommand{\kms}{km~s$^{-1}$}
\newcommand{\OI}{O~{\sc i}}

\newcommand{\SiII}{Si~{\sc ii}}
\newcommand{\CaII}{Ca~{\sc ii}}
\newcommand{\FeII}{Fe~{\sc ii}}
\newcommand{\FeIII}{Fe~{\sc iii}}
\newcommand{\TiII}{Ti~{\sc ii}}

\newcommand{\Nifs}{$^{56}$Ni}
\newcommand{\Mej}{$M_{\rm ej}$}
\newcommand{\KE}{$E_{\rm K}$}

\def\gsim{\mathrel{\rlap{\lower 4pt \hbox{\hskip 1pt $\sim$}}\raise 1pt \hbox {$>$}}}
\def\lsim{\mathrel{\rlap{\lower 4pt \hbox{\hskip 1pt $\sim$}}\raise 1pt \hbox {$<$}}}

\def\simlt{\mathrel{\hbox{\rlap{\hbox{\lower4pt\hbox{$\sim$}}}\hbox{$<$}}}}
\def\simgt{\mathrel{\hbox{\rlap{\hbox{\lower4pt\hbox{$\sim$}}}\hbox{$>$}}}}

\begin{document}

\title{A neutron-star-driven X-ray flash associated with supernova SN\,2006aj}

\author{%
Paolo A. Mazzali
    \affiliation[1]{Max-Planck Institut f\"ur Astrophysik,
     Karl-Schwarzschild Str. 1, D-85748 Garching, Germany\vspace{0.05in}}$^{,}$
    \affiliationmark[3]$^{,}$
    \affiliationmark[4]$^{,}$
    \affiliationmark[5]$^{,}$
    \affiliationmark[6],
Jinsong Deng
    \affiliation[2]{National Astronomical Observatories, CAS, 20A Datun Road,
     Chaoyang District, Beijing 100012, China\vspace{0.05in}}$^{,}$
    \affiliation[3]{Department of Astronomy, School of Science,
     University of Tokyo, Bunkyo-ku, Tokyo 113-0033, Japan\vspace{0.05in}}$^{,}$
    \affiliation[4]{Research Center for the Early Universe, School of Science,
     University of Tokyo, Bunkyo-ku, Tokyo 113-0033, Japan\vspace{0.05in}}$^{,}$
    \affiliationmark[6],
Ken'ichi Nomoto
    \affiliationmark[3]$^{,}$
    \affiliationmark[4]$^{,}$
    \affiliationmark[6],
Daniel N. Sauer
    \affiliationmark[5]$^{,}$
    \affiliationmark[6],
Elena Pian
    \affiliation[5]{Istituto Nazionale di Astrofisica-OATs, Via Tiepolo 11,
     I-34131 Trieste, Italy\vspace{0.05in}}$^{,}$
    \affiliation[6]{Kavli Institute for Theoretical Physics,
         University of California, Santa Barbara, CA 93106-4030, USA\vspace{0.05in}},
Nozomu Tominaga
    \affiliationmark[3]$^{,}$
    \affiliationmark[6],
Masaomi Tanaka
    \affiliationmark[3],
Keiichi Maeda
    \affiliationmark[6]$^{,}$
    \affiliation[7]{Department of Earth Science and Astronomy,
     College of Arts and Sciences, University of Tokyo, Komaba 3-8-1, Meguro-ku,
     Tokyo 153-8902, Japan\vspace{0.05in}},
Alexei V. Filippenko
    \affiliation[8]{Department of Astronomy, University of California,
     Berkeley, CA 94720-3411, USA\vspace{0.05in}}
}
\date{}{}
\headertitle{SN2006aj}
\mainauthor{Mazzali~\textit{et al.}}

%%%%%%%%%%%%%%%%%%%%%%%%%%%%%%%%%%%%%%%%%%%%%%%%%%%%%%%%%%%%%%%%%%%%%%%%%%

\summary{\small\bf

Supernovae connected with long-duration gamma-ray bursts
(GRBs)\cite{gal98,sta03,mal04} are hyper-energetic explosions
resulting from the collapse of very massive stars ($\sim 40
\Msun$, where $\Msun$ is the mass of the Sun) stripped of their
outer hydrogen and helium
envelopes\cite{iwa98,maz03,deng05,maz06}. A very massive
progenitor, collapsing to a black hole, was thought to be a
requirement for the launch of a GRB\cite{McF&W99}. Here we report
the results of modelling the spectra and light curve of
SN\,2006aj\cite{pian06}, which demonstrate that the supernova had
a much smaller explosion energy and ejected much less mass than
the other GRB-supernovae, suggesting that it was produced by a
star whose initial mass was only $\sim 20 \Msun$. A star of this
mass is expected to form a neutron star rather than a black hole
when its core collapses. The smaller explosion energy of SN~2006aj
is matched by the weakness and softness\cite{camp06} of GRB~060218
(an X-ray flash), and the weakness of the radio flux of the
supernova\cite{sod06}. Our results indicate that the supernova-GRB
connection extends to a much broader range of stellar masses than
previously thought, possibly involving different physical
mechanisms: a `collapsar'\cite{McF&W99} for the more massive stars
collapsing to a black hole, and magnetic activity of the nascent
neutron star\cite{tho04} for the less massive stars. } \maketitle

%%%%%%%%%%%%%%%%%%%%%%%%%%%%%%%%%%%%%%%%%%%%%%%%%%%%%%%%%%%%%%%%%%%

Like all other GRB-supernovae, SN~2006aj is of type
Ic\cite{pian06}. Its spectra resemble those of the dim,
broad-lined, non-GRB supernova SN~2002ap\cite{maz02,fol03}.
However, SN\,2006aj shows surprisingly weak oxygen lines for a
type Ic supernova. For a comparison of the spectrum of SN\,2006aj
to those of SN\,2002ap and of the GRB-supernova 1998bw, see
Supplementary Information.

To reproduce the spectrum of SN~2006aj\cite{pian06} we started
from the model that was used for SN~2002ap\cite{maz02}, but to
improve the spectral fits we reduced the masses of both oxygen and
calcium significantly, and decreased the ejected mass \Mej\ and
the kinetic energy \KE\ accordingly. The series of synthetic
spectra is shown in Figure 1.

A lack of oxygen lines in the spectrum suggests a small \Mej, but
it does not necessarily mean absence of oxygen in the ejecta. Our
model contains $\sim 1.3 \Msun$ of oxygen.  Oxygen is therefore
still the dominant element, but its abundance relative to other
(heavier) elements is much lower than in SN~2002ap or in the other
GRB-supernovae. Modelling also indicates that oxygen is confined
to high velocities (Fig. 1).  A shell of oxygen comprising $\sim
0.1 \Msun$ and expanding at velocities between 20,000 and
30,000\,\kms\ is detected, which may be the result of the episode
of interaction that was responsible for the early ultraviolet
brightening\cite{camp06}.

The spectroscopic results are confirmed by models of the light
curve. A synthetic light curve computed using the one-dimensional
density and chemical abundance structure obtained from the
spectral analysis reproduces the optical-infrared bolometric light
curve of SN~2006aj (Figure 2). For SN~2006aj we derive \Mej $
\approx 2 \Msun$ and \KE$ \approx 2 \times 10^{51}$\,erg. These
values are much smaller than those of the other GRB-supernova,
which typically have \Mej$ \approx 10 \Msun$ and \KE$ \approx 3
\times 10^{52}$\,erg\cite{iwa98,maz03,deng05,maz06}. The smaller
\KE\ and \Mej\ involved for SN~2006aj explain why the light curve
evolves more rapidly than that of SN~2002ap: the timescale of the
light curve depends in fact roughly on \Mej$^3$/\KE.\cite{arn82}
The supernova ejecta contain $0.21 \Msun$ of \Nifs, which is
responsible for the supernova luminosity.  About $0.02 \Msun$ of
this is located above 20,000\,\kms\ and causes the fast rise of
the light curve. The presence of \Nifs\ at high velocities is
unlikely to be the result of a spherically symmetric explosion. In
a realistic aspherical explosion, high-velocity \Nifs\ may be
copiously produced near the direction of the GRB jets\cite{mae02}.

Observations in the nebular phase, when the forbidden [\OI] 6300
and 6363~\AA\ lines should be strong in emission, will be needed
to determine more accurately the value of \Mej. Such observations,
to be performed starting August 2006, will also be useful to study
any possible asymmetry and the orientation of the supernova with
respect to the line of sight to the Earth, and thus to link the
supernova with the GRB\cite{mae02,maz05}.

The properties of both the supernova (small energy, small ejected
mass, low oxygen content) and those of the GRB (unusually soft and
long) seem to suggest that GRB~060218-SN~2006aj was not the same
type of event as the other GRB-supernovae known thus far.  The
radio properties of SN~2006aj were also intermediate between those
of the GRB-supernovae and of SN~2002ap\cite{sod06}.

One possibility is that the initial mass of the progenitor star
was significantly smaller than in the other GRB-supernovae, and
that the collapse/explosion generated less energy. A star of
zero-age main-sequence mass $\sim 20 - 25 \Msun$ would be at the
boundary between collapse to a black hole or to a neutron
star\cite{tom05}. If the star collapsed only to a neutron star,
more core material would be available to synthesize \Nifs. For
example, a star with $\sim 20 \Msun$ initially would have a
carbon-oxygen core of $\sim 3.3 \Msun$.\cite{tom05} If core
collapse left behind a neutron star of $\sim 1.4 \Msun$, $\sim 1.3
\Msun$ of oxygen and $\sim 0.6 \Msun$ of heavier elements
(including both intermediate-mass elements such as Si and Fe-group
elements) could be ejected in the supernova, consistent with our
results. Such a collapse is thought to give rise to an explosion
of \KE$\approx 10^{51}$\,erg\cite{nom94}, but there are
indications of a spread in both $E_K$ and the mass of \Nifs\
synthesized\cite{ham03}. Additionally, magnetar-type activity may
have been present, increasing the explosion energy\cite{tho04}.
Magnetic activity may also have caused the very long duration of
the $\gamma$-ray emission\cite{tho04} and the mixing-out of \Nifs\
required by the rapid rise of the light curve. It is also possible
that in this weaker explosion the fraction of energy channelled to
relativistic ejecta was smaller than in the classical
GRB-supernovae, giving rise to an X-ray flash (XRF).\cite{sod06}

Another case of a supernova associated with an XRF has been
reported\cite{fyn04}. The putative supernova, although poorly
observed, was also best consistent with the properties of
SN~2002ap\cite{tom04}. This may suggest that XRFs are associated
with less massive progenitor stars than those of canonical GRBs,
and that the two groups may be differentiated by the formation of
a magnetar\cite{nak98} or a black hole, respectively.  The
properties of both the GRB and the supernova may scale with the
mass of the progenitor\cite{nom05}. Still, the progenitor of
SN~2006aj had been thoroughly stripped of its H and He envelopes.
This is a general property of all GRB-supernovae known so far, and
possibly a requirement for the emission of a high energy
transient, which may be more easily achieved in a binary
system\cite{maz02,nom95,fry99}.

If the star was initially more massive ($\gsim 25 \Msun$), and it
collapsed directly to a black hole as in the more powerful
GRB-supernovae events, a number of questions arise. Why was the
energy of the explosion so small? Where did the large core mass
end up? Continuing accretion onto the black hole could explain the
missing mass. This may occur if the angular momentum of the core
was smaller than in the more energetic cases. Other more exotic
scenarios, such as merger models, may also work.

A case of a progenitor mass just exceeding the black hole limit
may be that of SN~2002ap. This SN may not have produced a magnetar
and an XRF, because it did not collapse to a neutron star but
rather to a black hole\cite{maz02}, yet at the same time the
energies involved in the collapse may have been too small to give
rise to a GRB.

In our scenario, some soft $\gamma$-ray repeaters energized by a
magnetar\cite{tho04,tho95} may be remnants of GRB~060218-like
events. Magnetars could thus generate a GRB at different times. As
they are born, when they have a very short spin period ($\sim 1$
ms), an XRF (or a soft GRB) may be produced as in
SN\,2006aj-GRB~060218. Later (after more than 1,000 years), when
their spin rate is much lower, they could produce short-hard GRBs
by a giant flare\cite{hur05}. Finally, if the progenitor star had
a massive companion in a close binary system, as may be required
for the outer envelope to be stripped and a long-duration GRB or
XRF to be produced\cite{fry99}, the system may evolve to a close
double-neutron-star system. When the two neutron stars finally
merge, a short-hard GRB may again be produced\cite{nar92}.

\begin{acknowledge}

We thank S. Kulkarni, C. Fryer, T. Janka, W. Hillebrandt, and C.
Kouveliotou for many stimulating discussions.  This work was
supported in part by the European Union, by the JSPS and MEXT in
Japan, and by the USA NSF.

\end{acknowledge}

\bigskip

\noindent
{\bf Author Information}

\noindent The authors declare no competing financial interest.
Correspondence should be addressed to P.A.M.
(mazzali@MPA-Garching.MPG.DE) or J.D. (jsdeng@bao.ac.cn).

%%%%%%%%%%%%%%%%%%%%%%%%%%%%%%%%%%%%%%%%%%%%%%%%%%%%%%%%%%%%%%%%%%%%%%%%%%%%%%%%%

\clearpage

\begin{figure}
\centerline{\psfig{file=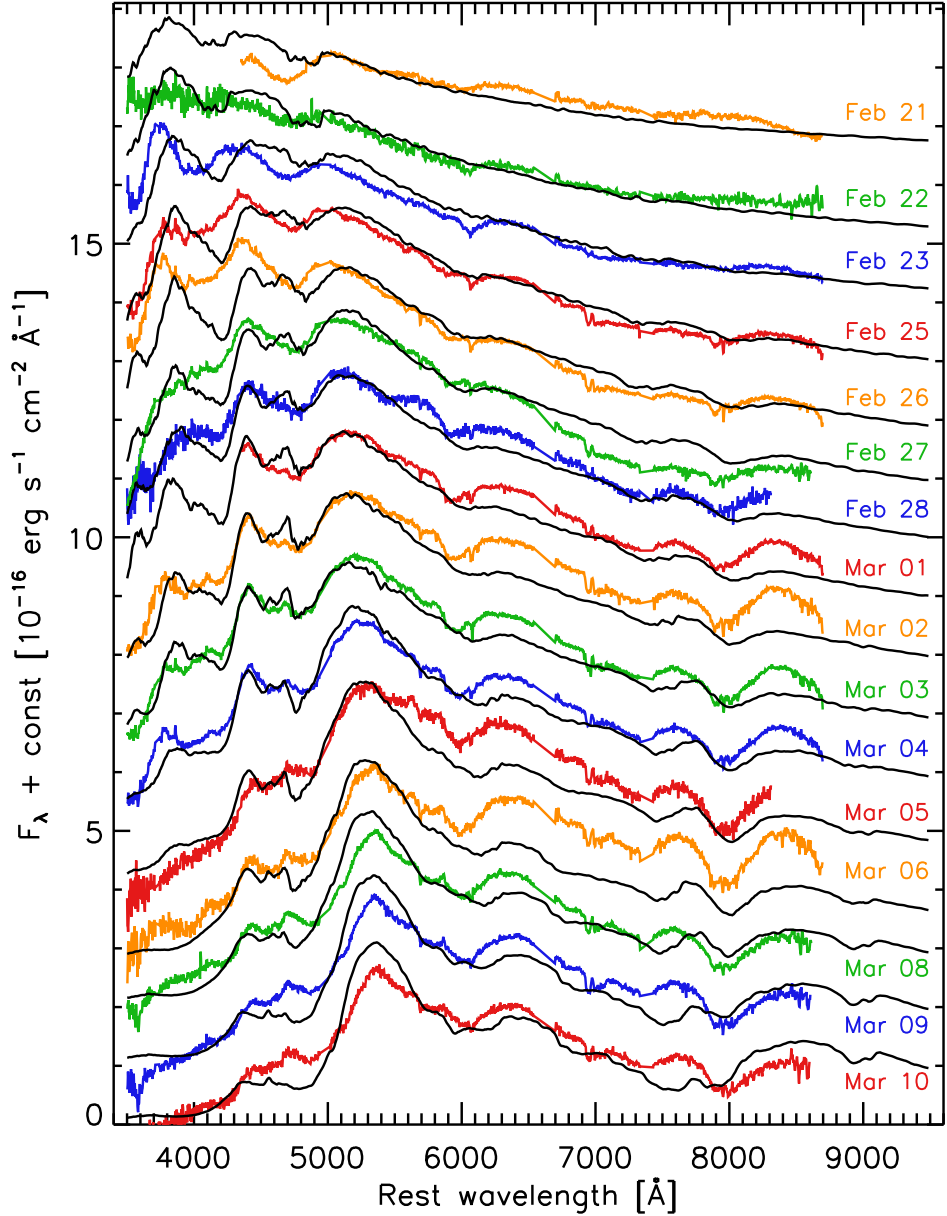,width=13.0cm}}

\caption{{\small {\bf Spectra of SN\,2006aj and synthetic fits.}
The observed spectra of SN\,2006aj (coloured traces) are
calibrated in the $V$ band, but elsewhere they my be
distorted\cite{pian06}, hence the poorer agreement in some of the
red parts. Also, the blue part is not reliable shortward of $\sim
4200$~\AA. The synthetic spectra (black traces) were computed
using our Monte Carlo spectrum synthesis code\cite{M00}. Because
of the spectroscopic and photometric similarity to
SN~2002ap\cite{fol03}, we used a similar model of the
explosion\cite{maz02},  but in order to improve the match we
reduced the masses of both oxygen and calcium significantly, and
decreased \Mej\ and \KE\ accordingly.  Our model  has \Mej$
\approx 2 \Msun$ and \KE $\approx 2 \times 10^{51}$\,erg. The
strongest features in the spectra are due to lines of \FeII,
\TiII, and in the later phases \CaII\ ($< 4500$~\AA), \FeIII\ and
\FeII\ (near  5000~\AA), \SiII\ (near 6000~\AA), \OI\ (near
7500~\AA), and \CaII\ (near 8000~\AA). The \OI\ and \CaII\ lines
become stronger at more advanced epochs, and are conspicuous
because they  form at a roughly constant wavelength, corresponding
to a velocity ($\sim 25,000$\,\kms) higher than that of other
lines. This indicates the presence of a shell of material,
dominated by oxygen, at velocities between about 20,000 and
25,000\,\kms. This high-velocity material may result from the
piling up of circumstellar material on the expanding ejecta. We
modelled the spectrum by adding a small amount of mass ($\sim 0.10
\Msun$) at $20,000 \lsim v \lsim 30,000$\,\kms. This results in an
increased \KE ($\sim 2.5 \times 10^{51}$\,erg). The fact that the
high-velocity material is mostly oxygen seems to confirm that both
the outer SN ejecta and the stellar wind were dominated by oxygen,
and that the progenitor star was an early-type WR star.}}{}
\label{fig:SpModels}
\end{figure}

\clearpage

\begin{figure}
\centerline{\psfig{file=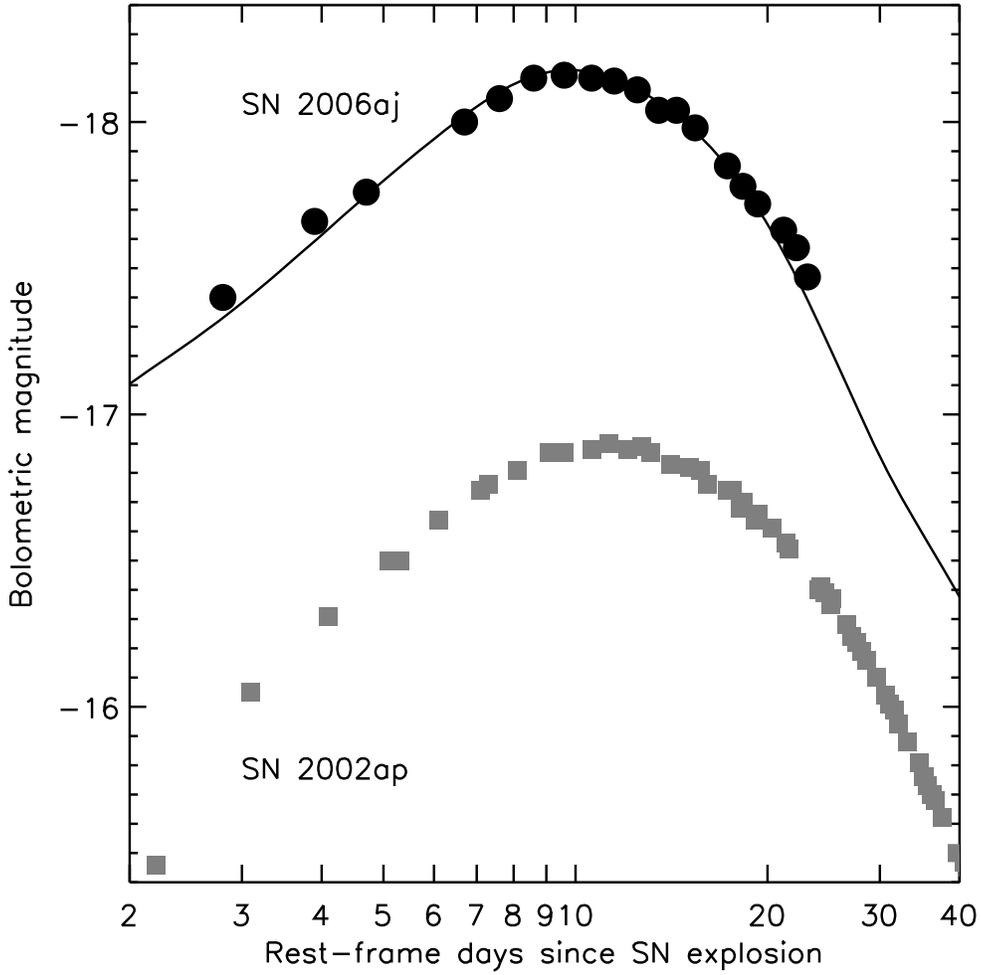,width=13.cm}}
\bigskip
\bigskip
\caption{{\small {\bf The light curve of SN 2006aj.} The
bolometric light curve of SN 2006aj (circles) is compared with the
model light curve (solid line), and with the bolometric light
curve of SN 2002ap (squares). A supernova light curve is powered
by $\gamma$-rays released in radioactive decays of freshly
synthesized unstable $^{56}$Ni to $^{56}$Co and hence to stable
$^{56}$Fe. The $\gamma$-rays deposit in the dense ejecta, giving
rise to a flux of optical photons. The light curve rises at first
as the diffusion time of photons decreases as the ejecta expand. A
maximum is reached when the escaping photon luminosity
approximately equals the deposited energy\cite{arn82}. The light
curve then declines as the density becomes low enough to allow
significant $\gamma$-ray escape. The more massive the supernova
ejecta and the smaller their kinetic energy, the more difficult it
is for photons to escape, which means that the light curve reaches
maximum later and has a broader peak. The bolometric light curves
were constructed by integrating the optical and near-infrared
fluxes (for SN 2006aj, optical photometry obtained with European
Southern Observatory's (ESO) Very Large Telescope (VLT) and
near-infrared photometry reported in the Gamma-Ray Burst
Coordinates Network (GCN) were used), after correcting for the
host-galaxy distance/redshift and the reddening toward the
supernova -- for SN 2006aj, 143 Mpc, $z=0.0335$, and
$E(B-V)=0.126$ mag\cite{pian06}]. The model light curve is
synthesized from the one-dimensional density and chemical
abundance structure of the best-fitting spectral models. It
corresponds to $\sim 2 \Msun$ ejecta expanding with a kinetic
energy of $\sim 2\times 10^{51}$ erg, having in total $\sim 0.2
\Msun$ of \Nifs. The small amount of mass and energy added by the
inclusion of the outer oxygen shell (see Fig. 1) have a very
limited impact on the light curve since the mass is located at low
density and has low optical depth. The explosion of SN 2006aj is
assumed to coincide in time with the GRB.}}{} \label{fig:LCurve}
\end{figure}

\clearpage

\end{document}